# Formation of Regular Satellites from Ancient Massive Rings in the Solar System.

Aurélien CRIDA[1*], Sébastien CHARNOZ[2,3]

**When a planetary tidal disk - like Saturn's rings - spreads beyond the Roche radius (inside which planetary tides prevent aggregation), satellites form and migrate away. Here, we show that most regular satellites in the Solar System probably formed this way. According to our analytical model, when the spreading is slow, a retinue of satellites appears with masses increasing with distance to the Roche radius, in excellent agreement with Saturn's, Uranus', and Neptune's satellites systems. This suggests that Uranus and Neptune used to have massive rings that disappeared to give birth to most of their regular satellites. When the spreading is fast, only one large satellite forms, as was the case for Pluto and Earth. This conceptually bridges the gap between terrestrial and giant planets' systems.**

Satellites are generally thought to form concurrently with a giant planet, in a large circum-planetary gaseous disk where there is inflow of solids. Two competing models exist in the literature *(1,2;3,4)*, in which solids aggregate to form satellites that can migrate in the gas (and possibly be lost) before the gas dissipates. These models have their pros and cons, but none can explain the surprising orbital architecture of Saturn's, Uranus' and Neptune's satellite systems, where the smallest bodies accumulate at a distance from the planet that is twice its radius (the Roche radius), and their masses increase with distance starting from this point (Fig. 1a). Moreover, in the frame of a circum-planetary gas disk, Uranus' satellites should orbit in the ecliptic plane, and not the equatorial plane of the tilted planet *(5)*. Also, Uranus and Neptune might be too light to have retained a massive enough gaseous disk *(6)*. All together this leads us to suggest that an alternative model is needed, to explain at least the origin of the giant planets' innermost satellites.

Here, we consider a disk of solid material around a planet, similar to Saturn's rings, wherein planetary tides prevent aggregation within the Roche radius $r_R$ *(7)* (Supplementary Material 1). It is known that such a tidal disk will spread *(8)*. Thus, the normalized disk lifetime can be defined by $\tau_{disk} = M_{disk} / FT_R$, where $F$ is the mass flow through $r_R$, $T_R$ is the orbital period at $r_R$ and $M_{disk}=\pi\Sigma r_R^2$ is the disk's mass ($\Sigma$ being the surface density). Using a prescription for the viscosity based on self-gravity and mutual collisions *(9)*, one finds

$$\tau_{disk} = 0.0425 / D^2 , \qquad \text{Eq.(1)}$$

where $D = M_{disk} / M_p$ and $M_p$ is the planet's mass (SM 2.2.1).

---

[1] Laboratoire Lagrange (UMR 7293),
Université Nice Sophia-antipolis / CNRS / Observatoire de la Côte d'Azur ; BP4229, 06304 Nice cedex 4, FRANCE

* To whom correspondence should be addressed. E-mail: `crida@oca.eu` .

[2] Laboratoire AIM, Université Paris Diderot / CEA/ IRFU-SAp ;    91191 Gif-sur-Yvette Cedex, FRANCE

[3] Institut Universitaire de France ;    103 Bd Saint Michel, 75005 Paris, FRANCE

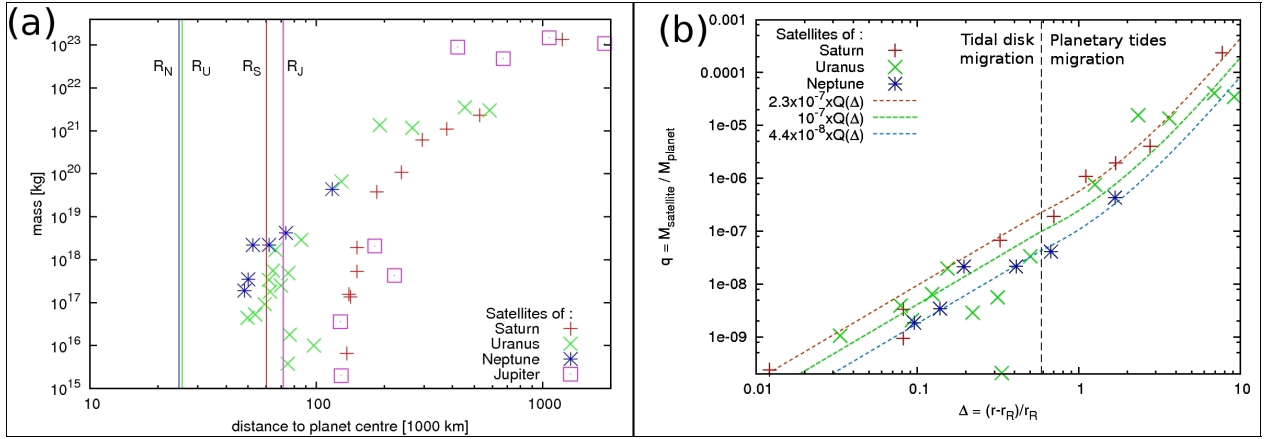

**Fig. 1**. Distribution of the regular satellites of the giant planets. Saturn: 9 satellites from Pandora to Titan. Uranus: (a) 18 from Cordelia to Oberon (b) 14 satellites, from Bianca to Oberon (except Cupid and Mab, out of scale). Neptune: Naiad, Thalassa, Despina, Galatea, Larissa and Proteus. Jupiter: Metis, Adrastea, Amalthea, Thebe, Io, Europa, Ganymede and Callisto.
(a) Mass as a function of the orbital radius. The four systems don't extend all the way down to the planetary radius (vertical line), but a pile-up of small satellites is observed at a specific distance (the Roche limit, $r_R$). The mass increases from zero with the distance to $r_R$, not to the center of the planets.
(b) Satellite-to-planet mass ratio q as a function of $\Delta=(r-r_R)/r_R$. The Roche radius for each planet is taken consistently with the mean density of satellites, or with the orbit of the closest one (SM 1). For Saturn, Uranus and Neptune, $r_R$=140 000 km, 57 300 km, 44 000 km respectively. Short dashed curves: our model for the pyramidal regime $Q(\Delta)$ (Eq.(S25), SM 6.3) : for $\Delta<\Delta_{2:1}$, $q \propto \Delta^{9/5}$ ; for $\Delta>\Delta_{2:1}$, $q \propto (\Delta+1)^{3.9}$, where $\Delta_{2:1}$=0.59 is marked by the vertical long dashed line. Jupiter's system does not fit well and is not shown (SM 7.3).

As material migrates beyond $r_R$, new moons form *(10, 11)*. They are then repelled by the tidal disk through resonant angular momentum exchange, and migrate outward as they grow. A satellite of mass M orbiting outside a tidal disk feels a positive gravitational torque *(12)*:

$$\Gamma = (32\pi^2/27)\ q^2\ \Sigma\ r_R^4\ T_R^{-2}\ \Delta^{-3}, \qquad \text{Eq.(2)}$$

where $q = M/M_p$, $\Delta = (r-r_R)/r_R$, and r is the orbital radius.
Thus, it migrates outwards at a rate (SM 2.1)

$$d\Delta/dt = (2^5/3^3)\ q\ D\ T_R^{-1}\ \Delta^{-3}. \qquad \text{Eq.(3)}$$

The migration rate increases with mass-ratio q and decreases with distance $\Delta$. Based on the restricted three-body problem, we assume that a satellite accretes everything within 2 Hill radii, $r_H$, from its orbit *(13, 14)* ($r_H=r(q/3)^{1/3}$). Thus, as a tidal disk spreads, a competition takes place between accretion and migration. We assume that the satellites don't perturb each other's orbit or the disk (SM 10), that $\Delta \ll 1$, and that D and $\tau_{disk}$ (thus F and $\Sigma$) are constant. We find from our analytical model (SM) that moon accretion proceeds in three steps, corresponding to three different regimes of accretion.

When the disk starts spreading, a single moon forms at the disk's edge (moon #1). As long as $\Delta<2r_H/r$, it will directly accrete the material flowing through $r_R$ and grow linearly with time, while migrating outwards. This is the "continuous regime" (Fig. 2a, SM 3). Integrating Eq.(3) with M=Ft, one finds that the condition $\Delta<2r_H/r$ holds until

$$\Delta = \Delta_c = (3/\tau_{disk})^{1/2} \qquad \text{Eq.(4)}$$
$$q = q_c = \sim 2\tau_{disk}^{-3/2}$$

Using Eq(1), one finds $\Delta_c$ = 8.4 D, occurring after ~10 orbits (SM 3.1).

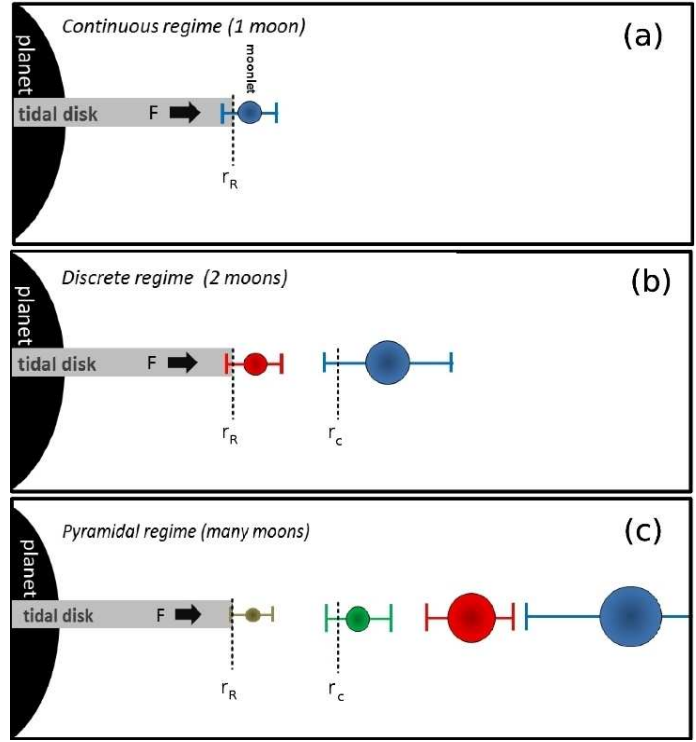

**Fig. 2.** Sketches of the three accretion regimes, where the accretion regions are defined as $\pm 2r_H$ around a moon's location. The tidal disk is in grey with F standing for the mass flow at the edge. The first moon to form is in blue, the second in red and the third in green.

(a) Continuous regime : only one moon is present, directly fed by the disk's mass flow ;

(b) Discrete regime : when the first moon has $r>r_c=r_R(\Delta_c+1)$, a new moonlet forms (in red). The first moon (blue) continues to grow by accreting these moonlets (red), which are fed by the disk and have masses $\leq M_c$ ;

(c) Pyramidal regime : when the first moon has $r-2r_H > r_c$, several moons can form and grow up to $M \geq M_c$. Moons with $r-2r_H > r_c$ grow through merging events between moons of similar masses.

Then, a second moon forms at $r_R$ (moon #2). Moon #2 migrates away rapidly, approaches moon #1 and is caught by it. Another moon forms, which is also eventually accreted by moon #1, and so on. The growth of moon #1 thus proceeds at the same average rate as before, but step by step through the accretion of moonlets : this is the "discrete regime" (Fig. 2b, SM 4). When $\Delta > \Delta_c + 2r_H/r$, moon #1 is too far away to be able to accrete a newly formed moon before this new moon leaves the continuous regime (SM 4.1). This corresponds to

$$\Delta = \Delta_d = \sim 3.1 \times \Delta_c \qquad \text{Eq.(5)}$$

and $q = q_d \sim 20\, \tau_{disk}^{-3/2} \sim 2200\, D^3$ as shown in SM 4. Eq.(1) implies this occurs after ~100 orbits. Also, $q_d < 0.1\, D$ provided $D < 6.7 \times 10^{-3}$. This is always the case around giant planets (see below and SM 7), and it justifies the assumption that D and $\tau_{disk}$ are constant. After $\Delta_d$ is reached, a third moon appears in the system, and the discrete regime ends.

As moons of fixed mass (produced by the discrete regime) appear successively at a given radius, they migrate outwards with decreasing rate, and hence their mutual distance decreases; eventually, they merge. Therefore, moons of double mass are periodically formed, migrate outwards, merge again, and so on. Assuming that the satellites don't perturb each other's orbit, mergers occur hierarchically – this is the "pyramidal regime" (SM 6). The moons' masses increase with distance, and an ordered orbital architecture settles. In the region $r<r_{2:1}=2^{2/3}r_R$, the migration is controlled by the disk's torque (Eq.(3)); then the mass-distance relation follows $M \propto \Delta^{1.8}$ and the number density of moons is proportional to $1/\Delta$ (SM 6.1). As a consequence, just outside $r_R$, an accumulation of small moons is expected, consistent with observations (Fig 1.a). Beyond $r_{2:1}$, the migration is controlled by the planet's tides and $M \propto r^{3.9}$ (SM 6.2).

This specific architecture is a testable observational signature of this process. A comparison with today's giant planets systems of regular moons reveals a very good match for Saturn, Uranus and Neptune (Fig. 1b). Neptune's inner moons (blue stars) match our model (blue dotted line) very well, except for Despina, whose mass is under predicted by a factor of 3. Uranus' satellites are somewhat more scattered but the system globally follows our model on the two sides of $r_{2:1}$ (SM 7.4), and the large number of satellites outside $r_R$ appears to be a natural byproduct of the pyramidal regime; however, the four main bodies are not ranked by mass. Saturn's case is especially remarkable because the 8 regular satellites from Pandora to Titan (considering Janus plus Epimetheus as one object) have masses that closely follow our model (red dashed line), through 4 orders of magnitude in distance and 6 in mass. Although unlikely, given our present understanding of Saturn's tides *(15)*, this result suggests that even Titan might have formed from the spreading of Saturn's initially massive rings. Titan being about 50 times more massive than Rhea could be understood in the frame of our model if the rings were initially very massive ($D\sim10^{-3}$, see below), favoring the formation of one dominant moon, which migrated fast outwards, taking most of the mass. Then, as the mass of the rings decreased, a standard pyramidal regime took place, giving birth to the other moons. However, Titan's tidal age (the time needed to reach its orbit through Saturn's tides) is estimated at about 10 billion years *(11)*; thus Titan's fit with our model may be a coincidence.

Jupiter's system isn't compatible with the pyramidal regime (SM 7.3), suggesting a different formation mechanism, even if the Laplace resonance may have erased the initial configuration. The system of the Galilean moons is satisfactorily explained by the circum-planetary disk models *(1-4)*. These models may also explain the formation of Saturn's moons Titan and Iapetus but not of the other satellites. In fact, the conditions inside Saturn's circum-planetary disc were very different. Jupiter opened up a gap in the proto-planetary nebula early in the history of the solar system, whereas Saturn, which formed later and is less massive, hardly did *(16)*.

Our model also predicts the conditions under which one moon is formed rather than numerous ones: surprisingly, the mass and distance of moon #1 at the end of the discrete regime depend on only one parameter, $\tau_{disk}$ (or D). Figure 3 shows the different regimes encountered during the recession of a moon in the $(\Delta,\tau_{disk})$ space (SM 8). When $\tau_{disk}$ is large, a moon that forms in the discrete regime reaches a low mass and small $\Delta$, as shown by Eq.(5). Thus, the pyramidal regime dominates, where many moons coexist and migrate away. For each of the Solar System's planets, the mass of the putative tidal disk is chosen to be about 1.5 times the mass of its current satellite system (SM 7) (Fig. 3). For all the giant planets, $D<2\times10^{-4}$, so that $\tau_{disk}>10^6$, making the pyramidal regime the final outcome, which explains the presence and the distribution of their numerous innermost regular satellites.

Conversely, when $\tau_{disk}$ is short, the continuous and discrete regimes are more efficient, and thus a massive satellite is built and migrates to large distances *(17)*. This applies to Pluto and the Earth, which have only one moon. Our model agrees well with N-body numerical simulations of the formation of Charon and the Moon *(10, 18-20)*, but neglects thermodynamical effects. This limitation is investigated in SM 7.1. In this case, where a single moon forms as the disk spreads, D varies strongly with time, making Eqs.(4,5) inaccurate. The system can still be solved provided F is constant, which is likely in a disk at thermodynamical equilibrium. The Earth's Moon probably finished its accretion in the discrete regime, allowing the possibility of a short-lived and low-mass companion moon. As the companion approached the proto-Moon, it may have been trapped in a horseshoe orbit, and later impacted the proto-Moon, as was recently suggested in order to explain the Moon's highlands *(21)*.

These results strongly suggest that, like the Moon and Charon, most regular satellites of Saturn, Uranus, and Neptune formed from the spreading of a tidal disk. Maybe only the giant planets' most massive regular satellites (the Galilean moons, Titan and Iapetus) formed directly from the planet's sub-nebula *(1-4)*. Many models have been proposed for the formation of the giant planets' massive rings. Rings could be remnants of disrupted satellites [either by an impact *(22, 23)* or by tides *(24)*], an explanation that is favored for Saturn, or remnants from tidally disrupted comets *(23, 25)*, which is more likely for Uranus and Neptune *(23)*. Uranus' and Neptune's large inclinations indicate that giant impacts capable of generating massive rings were common on the ice giants during the formation of the Solar System *(26)*. However, the many uncertainties in the initial conditions of giant planet formation make it hard to reach a conclusion on how massive rings formed. The way Uranus' and Neptune's massive rings disappeared is also an open question. Viscous spreading by itself would not be efficient enough to make the rings disappear because it becomes increasingly inefficient when the disk loses mass *(8)*. Atmospheric gas drag *(22, 27)*, the Yarkvosky effect *(28)*, or slow grinding of the rings through meteoritic bombardment *(22)* could all be invoked, but their role is still not well understood. However, our model is compatible with existing ring formation scenarios, and thus the most common mechanism of satellite formation is likely to be the spreading of a tidal disk surrounding a planet, terrestrial or giant, beside other processes *(1-4)*. The structure of the satellite system then depends only on the disk's lifetime.

**Acknowledgments:** This work was partially funded by the French Programme National de Planétologie.
S. Charnoz was supported by a "Labex UnivEarths" grant of Université Paris Diderot and by CEA/IRFU/SAP.
We thank A. Morbidelli, D. Nesvorny, and N. Murdoch for comments and help in the writing of this manuscript.

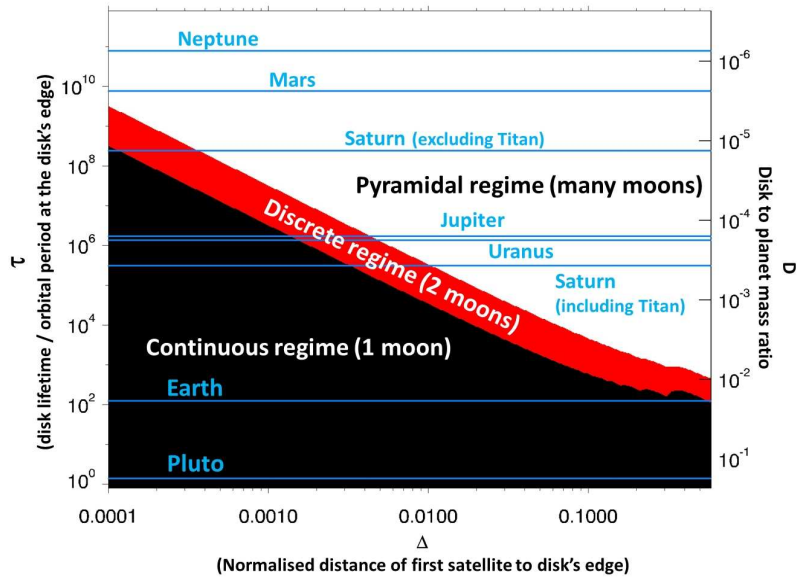

**Fig. 3.** Zones of the three regimes in the $\Delta - \tau_{disk}$ space obtained through numerical integration of the exact disk's torque (SM8). The right vertical axis displays the corresponding D to $\tau_{disk}$ through Eq.(1). When a disk spreads and forms satellites, a satellite follows a horizontal line ($\Delta$ increases and $\tau_{disk}$ is constant), from left to right. First, the satellite appears in the continuous regime (black region) where it is fed directly from the disk, while migrating away. Then it enters the discrete regime (red region) where it grows by regularly accreting new moons appearing at the disk's edge, in the black region. Finally, it leaves the discrete regime and many satellites may form and accrete all together – this is the pyramidal regime (white region). The boundaries between the regions follow exactly our analytical expressions for small $\Delta$ [Eqs.(4,5)]. Refined equations for the boundaries are provided in SM 9.
The horizontal lines show the paths that may have been followed around the Solar System's planets. The corresponding values of D were computed in SM7 and are for Jupiter, Saturn, Uranus, and Neptune $1.6\times10^{-4}$, $1.3\times10^{-5}$, $1.7\times10^{-4}$, $7.3\times10^{-7}$ respectively.

# Supplementary Material

# Formation of regular satellites from ancient massive rings in the Solar System.

Aurélien Crida & Sébastien Charnoz

## 1 The Roche radius of the giant planets

All the calculations and the model described in this paper are independent of the expression of the Roche radius $r_R$. However, in order to make Figure 1b and to test our model against the actual satellites systems, we need to know the value of $r_R$ around every giant planet.

The Roche radius around a planet of radius $R_p$ and density $\rho_p$, for material of density $\rho$, is given by (7):

$$r_R = 2.456 R_p \left(\frac{\rho_p}{\rho}\right)^{1/3} \quad \text{(S1)}$$

For Uranus, the average density (29) of the five largest regular satellites is 1627 kg.m$^{-3}$, so that Eq. (S1) gives $r_R = 57\,309$ km. This is between the orbits of the second innermost satellite of Uranus (Ophelia), and the third one (Bianca). We adopt this value.

For Jupiter, the average density (29) of the Galilean satellites is $\sim 2370$ kg.m$^{-3}$, leading to $r_R = 141\,000$ km.

In the case of Neptune, however, a similar calculation gives a Roche radius that would be $\sim 66\,000$ km, outside the orbit of four out of the six main regular satellites. We conclude that the density of these satellites must be ill determined. Thus, we take the Roche radius to be a little inside the orbit of Naiad (the innermost one), at 44 000 km. This corresponds to a density $\rho = 4240$ kg.m$^{-3}$ in Eq. (S1).

In the case of Saturn, the Roche radius is known to be between 138 000 and 141 000 km from the planet center, constrained by the presence of the rings. So, we take $r_R = 140\,000$ km. This corresponds to a density $\rho = 720$ kg.m$^{-3}$ in Eq. (S1). Note that the density of Janus is estimated to be 630 kg.m$^{-3}$.

## 2 Torque, migration, and disk evolution

### 2.1 Satellite-disk interaction

In the case where $\Delta = (r - r_R)/r_R \ll 1$, i.e., when a satellite is very close to the disk's outer edge, the gravitational torque exerted by the disk onto the satellitesimal is given at first order by (12):

$$\Gamma = \frac{8}{27} q^2 \Sigma r_R^7 \Omega_R^2 (r - r_R)^{-3} \quad \text{(S2)}$$

where $\Omega_R = \sqrt{GM_p/r_R^3}$ is the angular velocity at the Roche radius. In response to this torque, the orbital radius r changes following $\Gamma = M \frac{1}{2}\sqrt{GM_p/r}\left(\frac{dr}{dt}\right)$. Using $T_R = 2\pi/\Omega_R$, $M_{\rm disk} = \pi \Sigma r_R^2$, $D = M_{\rm disk}/M_p$, and $r \approx r_R$, this gives:

$$M\sqrt{GM_p/r}\frac{dr}{dt} = \frac{16}{27}q\Sigma r_R^2 \frac{GM_p}{r_R}\Delta^{-3}$$

$$\frac{1}{\sqrt{r_R\,r}}\frac{d(r-r_R)}{dt} = \frac{2^5}{3^3}qD\frac{\sqrt{GM_p}}{2\pi\,r_R^{3/2}}\Delta^{-3}$$

$$\frac{d\Delta}{dt} = \frac{2^5}{3^3}\,q\,D\,T_R^{-1}\Delta^{-3} \quad \text{(S3)}$$

### 2.2 Disk evolution

It is well-known that any viscous disk in Keplerian rotation spreads (e.g. (8)). We note $F$ the mass flow through $r_R$. Then, the life-time of the disk, normalized to the orbital period at $r_R$ reads:

$$\tau_{\rm disk} = \frac{M_{\rm disk}}{F\,T_R}\ . \quad \text{(S4)}$$

As we shall see below, this dimensionless parameter is central in the system.

From a hypothesis on the viscosity of the tidal disk, $F$ and $\tau_{\rm disk}$ are estimated as a function of the disk to planet mass ratio $D$ in the next subsection. However, in what follows, all expressions will be given both as a function of $\tau_{\rm disk}$ independently of any assumption on the viscosity, and as a function of $D$ under the hypothesis below.

### 2.2.1 Estimation of $F$ and $\tau_{\rm disk}$

The self-gravity and mutual collisions inside a disk cause a strong outward angular momentum flux. This can be described as an effective disk's viscosity. Using numerical simulations of self-gravitating disks, a semi-empirical expression for the effective viscosity has been derived *(9)*:

$$\nu \simeq 26\, r_H^{*\,5}\, G^2\, \Sigma^2\, /\, \Omega^3 \tag{S5}$$

where $r_H^*$ is a particle's Hill radius normalized by its diameter. Specifically, for $r = r_{\rm R}$, $r_H^* = 1.07275$.

The viscous spreading timescale, $T_\nu$, of a disk with effective viscosity $\nu$ and an outer radius $r_{\rm R}$ can be defined as $T_\nu \equiv r_{\rm R}^2/\nu$. Thus,

$$\begin{aligned}
\tau_{\rm disk} = T_\nu/T_{\rm R} &= \frac{r_{\rm R}^2 \Omega_{\rm R}}{2\pi \nu} \\
&= \frac{r_{\rm R}^2 \Omega_{\rm R}^4}{52\pi r_H^{*\,5} G^2 \Sigma^2} \\
&= \frac{1}{52\pi r_H^{*\,5}} \frac{G^2 M_p^2}{G^2 \Sigma^2 r_{\rm R}^4} \\
\tau_{\rm disk} &= \underbrace{\left(\frac{\pi}{52 r_H^{*\,5}}\right)}_{\sim 0.0425} D^{-2}
\end{aligned} \tag{S6}$$

With $\tau_{\rm disk}$ now expressed as a pure function of $D$, we can combine Eqs. (S4) and (S6) to obtain

$$F = \underbrace{\left(\frac{52 r_H^{*\,5}}{\pi}\right)}_{\sim 23.5} D^3 \frac{M_p}{T_{\rm R}} \tag{S7}$$

For massive planets with a small disk, the disk's normalized lifetime is large and the flow is small (e.g. Saturn's rings). The contrary stands for small planets surrounded by a massive tidal disk (e.g. the proto-lunar disk around the Earth).

## 3 Continuous regime

Let us consider that the material that crosses the Roche radius is immediately captured by an existing satellite outside. The mass of the latter grows continuously as $dM/dt = F$, the flow of material coming through the outer edge of the ring. We assume that $F$ is constant. This is valid on a time scale small with respect to the evolution of the disk. It is the case of Saturn's present rings, on the scale of the formation of the smaller moons up to Janus *(10)*. The migration rate of the satellite is given by Eq. (S3); with $M(t) = Ft$ and $\Delta(t=0) = 0$, this gives:

$$\Delta(t) = \frac{2^{3/2}}{3^{3/4}} D^{1/4} \left(\frac{M(t)}{M_p}\right)^{1/4} \left(\frac{t}{T_{\rm R}}\right)^{1/4} \propto t^{1/2}\, . \tag{S8}$$

This can also be written as a mass-distance relation using Eq. (S4):

$$q(t) = \left(\frac{\sqrt{3}}{2}\right)^3 \tau_{\rm disk}^{-1/2} \Delta(t)^2\, . \tag{S9}$$

### 3.1 End of the continuous regime

This regime applies only when the satellite is close enough to the disk that it can capture directly the material flowing through the Roche radius. This puts a constraint on the maximum distance of the satellite to the disk's edge: it must be smaller than $2r_H$, where $r_H = r_{\rm R}(q/3)^{1/3}$ is the Hill radius of the satellite *(13,14)*. Thus the condition of validity of the continuous regime is $\Delta(t) < 2(q/3)^{1/3}$. Using Eq. (S9), this translates into an upper limit of the satellite's mass or distance:

$$\begin{aligned}
\Delta &< 2\left(\frac{q}{3}\right)^{1/3} \\
\Leftrightarrow \Delta^3 &< \frac{2^3}{3}\left(\frac{\sqrt{3}}{2}\right)^3 \tau_{\rm disk}^{-1/2} \Delta^2 \\
\Leftrightarrow \Delta &< \sqrt{\frac{3}{\tau_{\rm disk}}} = \Delta_{\rm c} \approx 8.4\, D \tag{S10} \\
q &< \frac{3^{5/2}}{2^3} \tau_{\rm disk}^{-3/2} = q_{\rm c} \approx 222\, D^3 \tag{S11}
\end{aligned}$$

So, the continuous regime applies close to the outer edge of the rings, for $\Delta < \Delta_{\rm c}$ given by Eq. (S10). At first, material that leaves the rings form one single satellite, which grows continuously with $M = Ft$ and $\Delta$ given by Eqs. (S8) and (S9). As long as $q < q_{\rm c}$, this satellitesimal migrates so slowly that all the material that leaves the ring through the Roche radius is within its influence, and immediately accreted.

It is remarkable that the limits of the continuous regime, $\Delta_{\rm c}$ and $q_{\rm c}$, are simple expressions of one single parameter $\tau_{\rm disk}$ (or $D$ if one uses Eq. (S6) ).

Notice that $q_{\rm c}$ is much larger than the mass of the largest fragments that would form spontaneously from gravitational instabilities, which is $16\pi \xi^2 D^3$ *(30)*, where $\xi < 1$ (e.g. $\xi^2 = 0.1$). So, the continuous regime corresponds not only to the onset of the gravitational instabilities and the formation of clumps at $r_{\rm R}$, but leads to the formation of much larger bodies.

Since $q = Ft/M_p$, the time needed to reach the end of the continuous regime is $q_c M_p/F$, which, using Eq. (S7) is simply 9.5 orbits, independent of any other parameter.

# 4 Discrete regime

After the critical mass $q_c$ and distance $\Delta_c$ have been reached, the satellite is too far to accrete immediately the material that comes at $r_R$ from the tidal disk. So, it keeps migrating outwards at constant mass. A new satellitesimal forms, in the continuous regime. However, Eq. (S8) shows that this new satellitesimal migrates faster than the first one. Thus, it enters the first satellitesimal's sphere of influence, and is accreted before it reaches $\Delta_c$. The mass of the first satellite then increases. And this process repeats itself. In the end, the first satellite still follows $M = Ft$, but instead of accreting continuously new material directly from the tidal disk, it accretes regularly small satellitesimals. We call this the *discrete regime*, which is a natural prolongation of the previous *continuous regime*. This is illustrated by a numerical simulation in SM 5.

## 4.1 End of the discrete regime

The *discrete regime* is valid as long as

$$\begin{aligned}
\Delta &< 2\frac{r_H}{r_R} + \Delta_c \\
\Leftrightarrow \Delta &< \frac{2}{3^{1/3}}\frac{3^{1/2}}{2}\tau_{\text{disk}}^{-1/6}\Delta^{2/3} + \Delta_c \\
\Leftrightarrow \frac{\Delta}{\Delta_c} &< \left(\frac{\Delta}{\Delta_c}\right)^{2/3} + 1 \\
\Leftrightarrow 0 &< -1 + Z + Z^3 \quad,\text{with } Z = (\Delta_c/\Delta)^{1/3} \\
\Leftrightarrow Z_0 &< Z
\end{aligned}$$

where $Z_0$ is given by *(31)*:

$$Z_0 = \left(\frac{1+\sqrt{\frac{31}{27}}}{2}\right)^{1/3} + \left(\frac{1-\sqrt{\frac{31}{27}}}{2}\right)^{1/3} \approx 0.6823278$$

So, the domain of the discrete regime is

$$\Delta < \Delta_d = Z_0^{-3}\Delta_c \approx 3.15\Delta_c \approx \frac{5.45}{\sqrt{\tau_{\text{disk}}}} \approx 26.4\,D \quad (S12)$$

$$q < q_d = Z_0^{-6}q_c \approx 9.91 q_c \approx 19.3\,\tau_{\text{disk}}^{-3/2} \approx 2200\,D^3 \quad (S13)$$

The last satellitesimal accreted in the discrete regime has a mass $q_c$, while the first satellite is almost ten times more massive. So the growth of the mass of the first satellite proceeds through little steps.

Similarly, as $q_d = 9.91\,q_c$, the duration of the discrete regime is

$$t_d = 93.7\,T_R \quad (S14)$$

The discrete regime lasts about a hundred orbits at $r_R$, whatever value the other parameters have.

If $D$ is large, then $F$ is large and the migration is fast, so that in $\sim 10$ (resp. $\sim 100$) orbits, a satellite grows a lot and migrates far. If $D$ is small, $F$ is small and the migration is slow, so that in the end of these $\sim 10$ or $\sim 100$ orbits, one gets a small satellite close to the outer edge of the tidal disk.

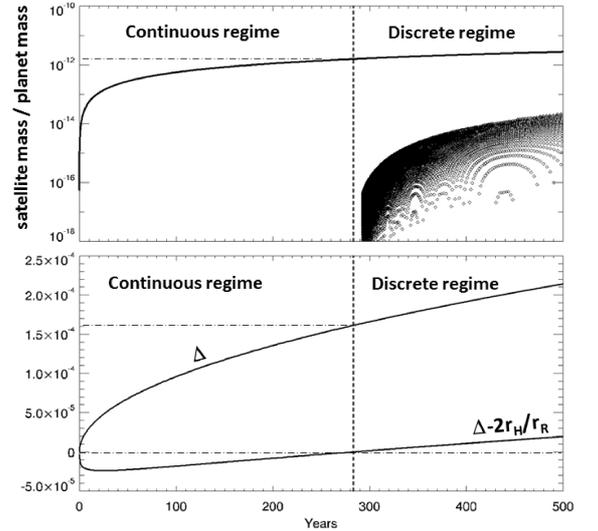

Figure S1: Numerical simulation of the continuous and discrete regimes.
**Top panel:** mass ratio of the satellite(s) to the planet as a function of time; the curve follows $q = Ft/M_p$; diamonds correspond to short-lived satellitesimals, quickly accreted by the larger one.
**Bottom panel:** $\Delta(t)$ (the curve matches Eq. (S8)), and the difference $\Delta - 2r_H/r$ for the first satellite formed.

# 5 An example

We have performed a numerical simulation, using a modified version of the code described in previous work *(10)*. In this simulation, we assume that the surface density of the disk is constant $\Sigma = 10$ kg.m$^{-2}$; this value is arbitrary (actually fairly low, of the order of that of Saturn's C ring). The disk is not evolved through viscosity, and its profile remains flat all the time, so that the torque on the satellite is exactly given by Eq. (S2). In addition, mass is input at the Roche radius at a rate $F = 10^5$ kg.s$^{-1}$. This flow is also arbitrary, and not given by Eq. (S7). Nonetheless, we can compute $\tau_{\text{disk}}$ using Eq. (S4). The mass input at $r_R$ is first turned into a satellite that migrates following Eq. (S3). Then, if this satellite is not further than 2 times its Hill radius from $r_R$, the mass input at $r_R$ is immediately acquired by it.

The simulation is run around a Saturn mass planet, with $r_R = 1.4 \times 10^8$ m, so that $T_R = 5.35 \times 10^4$ s. We have $\tau_{\text{disk}} = \pi\Sigma r_R^2/T_R = 1.15 \times 10^8$. Equations (S10) and (S11) give (using the expressions in $\tau_{\text{disk}}$, not in $D$):

$$\begin{aligned}
q_c &= 1.6 \times 10^{-12} \\
\Delta_c &= 1.6 \times 10^{-4} \\
t_c &= q_c M_p/F = 284 \text{ years.}
\end{aligned}$$

The result is shown in figure S1, and perfectly matches these predictions: the continuous regime takes place until a mass $q_c$ is reached after a time $t_c$. At this time, the satellite is at $\Delta_c$, and $\Delta - 2r_H/r$ becomes positive.

After, the discrete regime is entered, and many satellitesimals form before being accreted by the main satel-lite. They appear in the top panel of the figure as dots. As the first satellite migrates away, the satellitesimals have more space and time to grow and reach higher masses.

The transition between the continuous and discrete regime is very smooth, as the steps for the growth of the main satellite are so small that the curve still appears continuous. In fact, Eq. (S13) shows that the largest accreted satellitesimals are one order of magnitude smaller than the big one.

# 6  Pyramidal regime

In this section, we consider many satellites of constant mass, migrating outwards. Satellite-satellite interactions are here neglected, and we assume that the orbits are circular. First, we study migration under the influence of the tidal disk, then under the influence of the planetary tides. In both cases, the migration speed decreases with the distance. Therefore, the distance between two satellites decreases with time, as the faster inner one catches up with the slower outer one. Consequently, two satellites eventually have an encounter, and merge. The newly formed satellite is assumed to have the total mass of its two progenitors, and keeps migrating outwards.

Assuming that satellitesimals of mass $\delta m$ are produced every time interval $\delta t$ at a fixed radius $r$, then mergers between two of these first generation satellitesimals take place every $2\,\delta t$ at the same place $r_{m1}$. So, satellitesimals of mass $2\,\delta m$ are produced every $2\,\delta t$ at a fixed radius. Then, the process gets repeated, and these second generation satellitesimals will merge a little further, every $4\,\delta t$, leading to satellitesimals of mass $4\,\delta m$. And so on.

The regular production of satellitesimals of given mass, migrating outwards at a decreasing speed with distance, leads to a hierarchical or pyramidal merging history, like a genealogical tree. We call this the *pyramidal regime*. It is analytically described below. Recall that we saw in SM 4 that satellites of mass $M_d = q_d M_p$ are produced at $\Delta_d$ every $\delta t = M_d/F$. So, the pyramidal regime is the natural continuation of the continuous and discrete regimes.

## 6.1  Migration due to the tidal disk

For constant mass satellites, the solution of Eq. (S3) is:

$$\Delta(t) = \frac{2^{7/4}}{3^{3/4}} q^{1/4} D^{1/4} \left(\frac{t+t_0}{T_R}\right)^{1/4} \quad (S15)$$

where $t_0$ is an integration constant, determined by $\Delta(t=0) \equiv \Delta_0$.

Take two satellitesimals of mass $\delta m$, numbered 1 and 2, with number 1 born at time $t=0$ and number 2 at $t=\delta t$, at the location $\Delta_0$. Under the assumption $(t+t_0) \gg \delta t$, Eq. (S15) gives:

$$\Delta_1 - \Delta_2(t) = \left(\frac{q\,D}{2\times 3^3}\right)^{1/4} \frac{\delta t}{T_R} \left(\frac{t+t_0}{T_R}\right)^{-3/4} \quad (S16)$$

Two satellitesimals merge when their orbits are separated by less than two mutual Hill radii $r_{Hm} = r_R \left(\frac{2\,\delta m}{3\,M_p}\right)^{1/3}$, that is when

$$\Delta_1 - \Delta_2 < 2\left(\frac{2}{3}q\right)^{1/3} .$$

Then the merger of the two satellites occurs when $t = t_{\text{merge}}$ such that

$$\left(\frac{t_{\text{merge}}+t_0}{T_R}\right)^{3/4} = \left(\frac{1}{2^{19}\,3^5 q}\right)^{1/12} D^{1/4} \left(\frac{\delta t}{T_R}\right) . \quad (S17)$$

The merger takes place at a distance:

$$\Delta_1(t_{\text{merge}}) \approx \Delta_2(t_{\text{merge}}) \equiv \Delta_{\text{merge}}$$

$$\Delta_{\text{merge}} = \frac{2^{11/9}}{3^{8/9}}\, D^{1/3} \left(\frac{\delta t}{T_R}\right)^{1/3} q^{2/9} \quad (S18)$$

Such a merger leads to the formation of a satellite of mass $2\delta m$ at distance $\Delta_{\text{merge}}$. It occurs every $2\delta t$, as it needs 2 satellites. Then, the same process repeats itself with the products of the merging events. The above equations apply, replacing $\delta m$ by $2\delta m$ and $\delta t$ by $2\delta t$. The distance of the merging, $\Delta_{\text{merge}}$, is proportional to $\delta t^{1/3}\delta m^{2/9}$, so it is multiplied by $2^{5/9}$. In fact, $\Delta_{\text{merge}} \propto \delta m^{5/9} \propto \delta t^{5/9}$. In this pyramidal regime, the many satellitesimals have their masses roughly proportional the distance to the power 9/5:

$$\delta m \propto \Delta^{9/5} . \quad (S19)$$

Also, the number density of satellites should decrease with $\Delta$. The position of the $n^{\text{th}}$ satellite is $\Delta_n \propto 2^{5n/9} \approx 1.47^n$. So $n \propto \log(\Delta)$ and $dn/d\Delta \propto 1/\Delta$.

## 6.2  Migration due to planetary tides

Assuming the satellite is beyond the synchronous orbit, the planetary tides lead to the following migration rate *(32)*:

$$\frac{dr}{dt} = \frac{3k_{2p}(GM_p)^{1/2}R_p^5}{Q_p}\, q\, r^{-11/2} \quad (S20)$$

with $k_{2p}$, $M_p$, $R_p$, and $Q_p$ are standing for the planet's tidal Love number, mass, equatorial radius, and dissipation factor. With $K = 39k_{2p}(GM_p)^{1/2}R_p^5/2Q_p$ assumed constant, the solution of this equation is:

$$r(t) = (Kq)^{2/13}(t+t_0)^{2/13} \quad (S21)$$

Again, we take two satellitesimals of mass $\delta m$, numbered 1 and 2, with number 1 born at time $t=0$ and number 2 at $t=\delta t$, at the location $r_{t0}$. Under the assumption $(t+t_0) \gg \delta t$, Eq. (S21) gives:

$$r_1 - r_2(t) = \frac{2}{13}(Kq)^{2/13}(t+t_0)^{-11/13}\delta t \quad (S22)$$

So, a merger takes place when

$$r_1 - r_2(t) = 2\left(\frac{2}{3}q\right)^{1/3} r(t)$$

$$\frac{2}{13}\frac{(Kq)^{2/13}}{(t+t_0)^{11/13}}\delta t = 2\left(\frac{2}{3}q\right)^{1/3}(Kq)^{2/13}(t+t_0)^{2/13}$$

$$\frac{\delta t}{13(t+t_0)} = \left(\frac{2}{3}q\right)^{1/3}$$

$$t+t_0 = \frac{3^{1/3}\delta t}{26\,q^{1/3}}$$

That is at

$$r_{\text{merge}} = (Kq)^{2/13}\frac{3^{2/39}}{26^{2/13}}\delta t^{2/13}q^{-2/39} \propto \delta t^{2/13}q^{4/39} \tag{S23}$$

As $\delta t$ and $q$ are doubled every merging event, the many satellitesimals have their orbital radius roughly proportional the mass to the power $10/39$:

$$\delta m \propto r^{39/10} \;. \tag{S24}$$

Note this power-law is independent of the numerical coefficient, and in particular of the efficiency of the dissipation inside the planet $k_{2p}/Q_p$, provided the latter is independent of $r$ and constant with time. All the satellites that form in the pyramidal regime with migration dominated by the planetary tides will align on a straight line of slope 3.9 in a log(mass) – log(orbital radius) diagram. The orbital radius of the outer most satellite, though, is determined by $K$ and by the total time of evolution of the system. This sets the domain of application of our model.

### 6.3 Conclusion

Two different mass-distance relations are found. When the distance of a satellite to the tidal disk is such that no more inner Lindblad resonances with the satellite fall within $r_\mathrm{R}$, then the torque from the disk vanishes. This happens when the 2:1 inner Lindblad resonance falls at $r_\mathrm{R}$, and corresponds to $\Delta = \Delta_{2:1} \approx 0.587$. Beyond that, Eq. (S24) applies. For smaller $\Delta$, one should compare the two migration rates given by Eqs. (S3) and (S20). With $\bar{a} = r/r_\mathrm{R}$ and $r_\mathrm{R}$ given by Eq. (S1), one finds:

$$\frac{(d\bar{a}/dt)_\text{tides}}{(d\Delta/dt)_\text{disk}} = \underbrace{\left(\frac{3}{2}\right)^4 \pi\, 2.456^{-5}\left(\frac{\rho_p}{\rho}\right)^{-5/3}\left(\frac{k_{2p}}{Q_p}\right)}_{\sim 4\times 10^{-5}}\frac{1}{D}\frac{\Delta^3}{\bar{a}^4}$$

where we have taken $\rho_p/\rho \approx 1$ and $k_{2p}/Q_p = 2.3 \times 10^{-4}$ *(15)*.

Finally:

$$\frac{(d\bar{a}/dt)_\text{tides}}{(d\Delta/dt)_\text{disk}} < \begin{cases} 10^{-6}/D & \forall \Delta < \Delta_{2:1} \\ 10^{-7}/D & \forall \Delta < 0.17 \\ 10^{-9}/D & \forall \Delta < 0.03 \\ 0.7\, D^2 & \forall \Delta < \Delta_d \end{cases}$$

So, the torque from the planetary tides is negligible in most cases for $\Delta < \Delta_{2:1}$ (see estimates of $D$ around Solar System planets in SM 7: $D > 10^{-5}$ except for Neptune and Mars).

Note in passing that the torque from the planetary tides is always negligible for $\Delta < \Delta_d$, which validates *a posteriori* the use of the disk torque in the calculations relative to the continuous and discrete regimes.

Satellites of same mass should be regularly delivered at $\Delta_{2:1}$ by the pyramidal regime dominated by the torque from the tidal disk. Then, the pyramidal regime dominated by the planetary tides takes it over. In the end, satellites formed in the pyramidal regime have their masses such that:

$$q \propto \mathcal{Q}(\Delta) = \begin{cases} (\Delta/\Delta_{2:1})^{9/5} & \Delta < \Delta_{2:1} \\ \left(\frac{\Delta+1}{\Delta_{2:1}+1}\right)^{3.9} & \Delta > \Delta_{2:1} \end{cases} \tag{S25}$$

This relation is plot on Figure 1b, and fits quite well the satellites systems of Saturn, Uranus, and Neptune. For very low massive disks, the transition between the two power laws could take place at smaller $\Delta$, but this would not change much the shape of the curve in the figure, as the transition is relatively smooth.

## 7 Application to Solar System bodies: Estimation of the tidal disks' initial masses and $\tau_\text{disk}$

How can we estimate the mass of the tidal disk, that may have given birth to the satellite population of the different planets? A good starting point is to consider the satellites systems as seen today. Indeed, several works have shown that when a tidal disk spreads, the mass implanted in the satellite(s) is comparable to the mass of the full tidal disk, and that some mass is lost onto the planet *(10,11,18,19)*. So, if today's satellites have formed from a tidal disk, the mass of the latter, $M_\text{disk}$ must have been comparable to the present mass of the satellites. Thus, we apply the following simple rule: we assume that $M_\text{disk} \simeq 1.5\, M_s$, where $M_s$ is the total mass of all satellites that we assume to be born from the tidal disk. $M_s$ must be estimated for each planet. From this and Eq. (S6), we can derive a value of $D$ and of $\tau_\text{disk}$ for these planets.

- **Earth:** $M_s = 7.3 \times 10^{22}$ kg, $D = 0.0183$, $\tau_\text{disk} = 125$.

- **Mars:** Although most researchers favor a capture origin for both Phobos and Deimos, it has been suggested *(33)* that Mars had a debris disk with mass $M_\text{disk} \approx 10^{18}$ kg, so that $D \approx 2.3 \times 10^{-6}$, and $\tau_\text{disk} \approx 7.7 \times 10^9$.

- **Jupiter:** $M_s = 2 \times 10^{23}$ kg, mass of all Galilean satellites, $D = 1.6 \times 10^{-4}$, $\tau_\text{disk} = 1.7 \times 10^6$.

- **Saturn:** $M_s = 5 \times 10^{21}$ kg mass of all satellites below Titan (as most researches favor the theory that Titan and Iapetus formed in Saturn's nebula),

$D = 1.3 \times 10^{-5}$, and $\tau_{\rm disk} = 2.4 \times 10^8$.
Taking $M_s = 1.4 \times 10^{23}$ kg (mass of all satellites up to and including Titan, which fits well in our model) gives $D = 3.7 \times 10^{-4}$, and $\tau_{\rm disk} = 3.1 \times 10^5$.

- **Uranus :** $M_s = 10^{22}$ kg, mass of all satellites up to (and including) Oberon, $D = 1.7 \times 10^{-4}$, and $\tau_{\rm disk} = 1.4 \times 10^6$.

- **Neptune :** $M_s = 5 \times 10^{19}$ kg, mass of all satellites apart from Triton (which is thought to be captured), $D = 7.3 \times 10^{-7}$, and $\tau_{\rm disk} = 7.9 \times 10^{10}$.

- **Pluto :** $M_s = 1.5 \times 10^{21}$ kg, Charon's mass, $D = 0.17$, and $\tau_{\rm disk} = 1.4$.

The horizontal lines on Figure 3 correspond to these numbers.

## 7.1 Discussion about massive disks : cases of the Earth and Pluto

### 7.1.1 Comparison with numerical simulation

Here, our model is compared with previous work, in which N-body simulations of the impact were made. No thermodynamics is at play in this case, so that the viscosity is by construction given by Eq. (S5), and $\tau_{\rm disk}$ by Eq. (S6). This may not be physical (see below), but allows an easy comparison.

For the Earth, with $\rho = \rho_{\rm Moon} = 3346$ kg.m$^{-3}$, we have $r_{\rm R} \approx 1.85 \times 10^7$ m, and $T_{\rm R} = 7$ h. So, we expect the disk to disappear in about 36.5 days, in correct agreement with N-body simulations suggesting about a month *(19)*.

In this frame, the Moon should have ended at $\Delta = \Delta_{2:1} = 0.587$, which is at $r = 1.58 r_{\rm R} = 2.9 \times 10^7$ m. At this distance, the orbital period around the Earth is 13h 50min. Thus, one could think that the Moon was below the synchronous orbit. However, for the Earth-Moon system to have the present total angular momentum, the rotation period of the Earth at this time needs to be 6h 6min. So, in our model, the Moon forms well beyond the synchronous orbit, and the Earth-Moon system has evolved through tidal effects to the present state.

Also, $\tau_{\rm disk} = 125$ gives $q_{\rm d} = 0.0140$, which is just a little bit more than the mass ratio of the Moon to the Earth (0.0122), and just a little bit less than the putative $D = 0.0183$; so, the Moon should have finished its accretion at the end of the discrete regime. This allows for the formation of satellitesimals of mass a less than a tenth of the Moon, which should be accreted by the Moon. However, it is possible that one of these satellitesimals is captured in a horseshoe orbit (it happened to Epimetheus with Janus). This supports the idea that a small companion to the Moon was accreted later, forming the lunar highlands *(21)*.

Concerning Pluto, we find that $\tau_{\rm disk} \approx 1.4$, which means a spreading time of the order of the orbital period at the disk's edge. Whereas shocking at first sight, this result is in agreement with SPH simulations of Charon formation, after an impact on the proto-Pluto. Indeed, in these simulations, only 10-15 hours after the impact, the disk is emptied or the proto-Charon is formed *(20)*. Assuming that the proto-Pluto was about Pluto's mass, this yield an orbital period at the disk's edge about 12 hours, which means that $\tau_{\rm disk}$ is about 1. Our computed value ($\sim 1.4$) is in the right order of magnitude, and is compatible with an emptying timescale comparable with the orbital period.

Our qualitative agreement with SPH simulations in these extreme conditions shows that we may still catch the important physics at play.

### 7.1.2 Rapidly evolving disks : first insight into thermodynamical effects

In the case of massive disks and small $\tau_{\rm disk}$, Eq. (S13) for $q_{\rm d}$ doesn't apply because $\Sigma$ and $D$ can't be considered as constant when $q_{\rm d} \approx D$. In addition, for massive disks, $\tau_{\rm disk}$ is not given by Eq. (S6) because the viscosity is probably not given by Eq. (S5) in the Moon-forming disk : the viscous heating would lead to its melting or vaporization.

A model of viscosity limited by the black-body cooling of the photosphere of the Moon-forming disk *(34)* makes the dissipation constant, hence the product $\Sigma \nu$ and the flow $F$. In this case, we still have $M(t) = Ft$. Thus, Eq. (S3) can still be integrated, with $D = D_0 - q$, $D_0$ being the initial disk to planet mass ratio. It gives :

$$q\sqrt{1 - \frac{2}{3}\frac{q}{D_0}} = \left(\frac{\sqrt{3}}{2}\right)^3 \tau_{\rm disk}^{-1/2} \Delta^2 \qquad (S26)$$

with $\tau_{\rm disk} = D_0 M_p / F T_{\rm R}$. The condition for the continuous regime turns into an implicit condition for the mass of the satellite :

$$q\left(1 - \frac{2}{3}\frac{q}{D_0}\right)^{3/2} < \frac{3^{5/2}}{2^3}\tau_{\rm disk}^{-3/2} = q_{\rm c} \qquad (S27)$$

The continuous regime can now be maintained with $q$ slightly larger than $q_{\rm c}$ (but note that here, $q_{\rm c}$ is not any more $222 D^3$). This is because in this model, the flow (accretion) is constant, while $\Sigma$ and $D$ (which govern migration) decrease with time. In fact, the left hand side of Eq. (S27) is always smaller than $3D_0/2^{7/2}$, so that the continuous regime is never left if $\tau_{\rm disk} < 2^{1/3} 3 D_0^{-2/3}$. This corresponds to a spreading time of one month for the Earth, and 6 days for Pluto. So, with a constant flow $F$ (which is justified by thermodynamical constraints), taking into account the fact that $D$ decreases with time, we find that Charon grew entirely in the continuous regime, while the Moon probably started the discrete regime. But the thermal and dynamical evolution of the Moon-forming disk is a very complex problem, far beyond the scope of this paper.

Of course, in such a high disk/planet mass ratio and on such short timescale, other approximations of our model are violated. More precisely : (a) the gravitational potential may be far from Keplerian ; (b) the computation of disk's torque assumes that all resonances have time to respond to the satellite's perturbation, which is not possible in one orbital period; (c) for such a value of $\tau_{\rm disk}$, $\Delta_c$ is not small, and the approximation $\Delta \ll 1$ is not valid (see SM 9 for the large $\Delta$ case).

## 7.2 About Mars

Assuming a disk of mass $10^{18}$ kg, we find that the spreading of this disk should have been dominated by the pyramidal regime. Indeed, $\tau_{\rm disk}$ is extremely high, so that $q_d \approx 10^{-14}$, which correspond to objects of $\sim 5 \times 10^9$ kg only. If Phobos and Deimos were to be formed this way, it means that a chain of satellites formed, of masses increasing with distance as described in SM 6. Phobos and Deimos would have been the two last and most massive ones.

Subsequent orbital evolution would have made all satellites inside the orbit of Phobos crash on Mars, because they were below the synchronous orbit. This scenario has been recently proposed *(33)*, in which a giant impact creates the disk around Mars from which Phobos and Deimos form. Our model shows that in this case, the pyramidal regime should prevail, explaining the many past Martian moons and the two presently surviving ones. Note that only a few moons would have had masses of the order of Phobos, the other ones being much smaller. As a consequence, we don't expect too many large oval craters on the surface of Mars created by this population.

Nonetheless, the origin of Phobos and Deimos is still a matter of debate. We tried here to apply our model to Mars, but it is difficult to reach a conclusion, and this issue is beyond the scope of this paper. In particular, it should be noted that Deimos may not have formed in the pyramidal regime because it can't be transported beyond Mars' synchronous orbit. Then, only Phobos may be the last member of a population of satellites formed from the spreading of a tidal disk, and Deimos may have been captured.

## 7.3 About Jupiter

We find that if Jupiter had been surrounded by a tidal disk of mass equivalent to its satellite system, it should have been dominated by the Pyramidal regime. This is consistent with the fact that Jupiter has 8 regular satellites, and not one. However, the system of the regular satellites of Jupiter is divided in 2 groups of 4 bodies: the Galilean satellites which orbit beyond $4 \times 10^8$ m and have masses above $10^{22}$ kg; Metis, Adrastea, Amalthea and Thebe, which are inside $2 \times 10^8$ m and are lighter than $2 \times 10^{18}$ kg. Among these two groups, the mass is not an increasing function of the radius. This is not in qualitative agreement with the Pyramidal regime.

The system is represented in Figure S2, which is similar to Figure 1b in the main text for the other giant planets. Note that Metis and Adrastea do not appear in this figure because they orbit inside the Roche radius as calculated in SM1; they could be denser than the aver-age of the Jovian moons, or have slightly migrated inward since they formed. Anyway, the fit with $\mathcal{Q}(\Delta)$ defined by Eq. (S25) is not good. So, it seems that the regular satellites of Jupiter didn't form from a tidal disk. Note that there is so far no explanation for the strong differences between these two groups of satellites. In particular, the mass-distance distribution and the composition of the smaller ones is still puzzling. In addition, the Laplace resonance couples the orbital evolution of Io, Europa and Ganymede, which erases the initial conditions. Thus, it is difficult to reach a conclusion about the origin of this system. Note that models indicate that the conditions inside Jupiter's circum-planetary disk were significantly different from that in Saturn's one. The four Galilean moons could have formed in such a circum-planetary disk (in terms of orbital architecture and composition), in contrast to that of Saturn, where only one Titan is expected to form *(1-4)*. Because it formed earlier, and quickly opened a gap in the circum-solar nebula *(16)*, Jupiter could be the only giant planet whose satellite system formed entirely inside its circum-planetary disk.

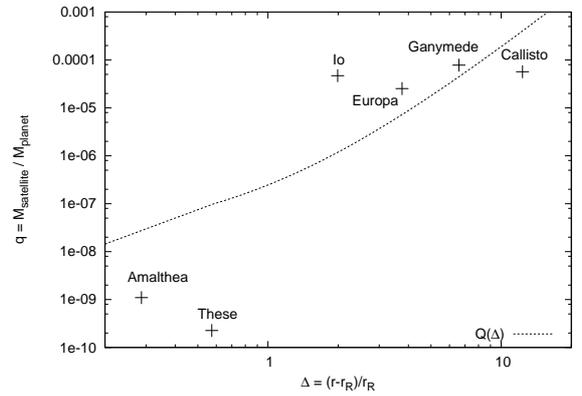

Figure S2: The satellites of Jupiter in the $q - \Delta$ plane, with $r_{\rm R} = 1.41 \times 10^8$ m (see SM 1 and 7.3). The dashed line follows Eq. (S25).

Also, it could well be that Jupiter simply never had a massive ring system.

## 7.4 About Uranus

Uranus has many small satellites close to its Roche radius. This is in agreement with the beginning of the pyramidal regime. Their masses globally follow $q \propto \Delta^{9/5}$, but some scatter is observed. Actually, it has been shown that this system is chaotically unstable *(35)*, but the dynamical instabilities occur on time scales much longer than that of the formation of this system ($\tau_{\rm disk}$ corresponds to $\sim 1600$ years). Our assumption that the satellites don't perturb each other's orbit is therefore valid for the duration of the formation of the satellites. The Uranian satellites probably formed with $q \propto \Delta^{9/5}$, and this distribution has been altered afterwards by chaotic motion, leading to the observed scatter. In fact, the potentially destructive chaos in the smallest moons of Uranus within times of the order of $4 - 100$ million years argues in favor of a younger formation for these bodies than for the planet itself *(35)*. At least the small moons of Uranus probably didn't form in a circum-planetary gaseous disc.

It should be noted that the tiny rings shepherded by satellites around Uranus shouldn't be considered as remnants of the original tidal disk. They are more likely remnants of the accretion process and of collisions between satellites.

## 7.5   About Neptune and Triton

Triton is the largest satellite of Neptune, but is an irregular satellite with a retrograde orbit. Therefore, is is thought to have been captured.

In the case of a capture through an impact, this event likely perturbed the orbits of the pre-existing satellites outside of 5 Neptune radii, leading to their possible ejection or merging with Triton *(36)*. So, the 6 regular satellites observed today could have escaped destruction, if they were already formed. It could also be that Neptune had more satellites in the past, before the capture of Triton, which prolonged the $q - \Delta$ relation observed in Fig. 1b. In this case, the disk mass estimated here is a lower bound.

One could also imagine that the formation of the regular satellites took place after Triton's capture. After all, the impact that led to the capture of Triton could also have produced a tidal disk; this idea is appealing, but considering Triton's angular momentum, this tidal disk would have been retrograde, which is not in agreement with the present prograde orbits. However, if Triton impacted Neptune, other bodies of similar size may well have done so at a later time, giving eventually birth to the regular satellites through a tidal disk.

A more recent model considers the capture of Triton from a binary object *(37)*. This event may have been primordial, and may have not destroyed the putative satellite system. In that case, the formation of the 6 regular satellites of Neptune could be primordial as well and not much perturbed, or easily posterior to Triton's capture.

In conclusion, the presence of Triton and its necessary capture doesn't contradict the idea that our model applies to Neptune's regular satellites.

## 8   Numerical exploration of all regimes using exact resonant torque

Since $\tau_{\text{disk}}$ was found to be the only free parameter, we investigate numerically how the mass of the first moon increases as a function of $\Delta$ and $\tau_{\text{disk}}$. For a given value of $\tau_{\text{disk}}$ (ranging from 0.1 to $10^8$) we solve simultaneously the following equations:

(a) the evolution of $r$, the distance of the satellite to the planet's due to the disk's torque *(11)*:

$$\frac{dr}{dt} = \frac{2r^{1/2}}{M(GM_p)^{1/2}} \sum_{m=2}^{+\infty} \Gamma_m \quad \text{(S28)}$$

(b) the evolution of the mass of the satellite

$$\frac{dM}{dt} = F \quad \text{(S29)}$$

$\Gamma_m$ is the torque applied by the disk onto the satellite by the $m-1:m$ resonance inside the disk *(38)*. We used an exact torque expression *(39)*, valid either for low and high $m$. At each time step, as $r$ increased, we checked whether the satellite was in the continuous, discrete, or pyramidal regime. The results are plotted in Figure 3. The discontinuous aspect of the frontiers between the different regimes, for high $\Delta$ (or low $\tau_{\text{disk}}$), is due to the discontinuous leaving of the resonances from the disk as the big satellite goes away: the 4:3 resonance, then the 3:2 and finally, the last one, the 2:1. When the 2:1 leaves the disk, then the satellite stops. This corresponds to $\Delta = \Delta_{2:1} = 0.587$.

## 9  Large $\Delta$

When the first moon is very far from the Roche radius, then the hypothesis $\Delta \ll 1$ does not hold anymore. To explore this regime, we look for a more precise expression of $\Delta_c$. When $\Delta$ is not small, the expression of the torque exerted by the disk on the outer satellite should be taken from *(39)*, their Eq. (47b). It gives:

$$\frac{d\Delta}{dt} = \frac{2}{3} f' \, q \, D \, \bar{a}^{9/2} \, T_R^{-1} \Delta^{-3} \qquad (S30)$$

where $f' \approx 2.5$. Not only the numerical coefficient is slightly different, but more importantly, a dependence in $\bar{a} = r/r_R = \Delta + 1$ appears, which makes the integration less easy. Using $q = Ft/M_p$, $d\bar{a} = d\Delta$, and separating the variables, one finds:

$$(\bar{a}-1)^3 \bar{a}^{-9/2} \, d\bar{a} = \frac{2f'}{3T_R} D \frac{Ft}{M_p} \, dt \; . \qquad (S31)$$

The right hand term integrates into:

$$\frac{f' \tau_{\text{disk}}}{3} q^2 \; .$$

With the condition that when $t=0$, $q=0$ and $\bar{a}=1$, the left hand term of Eq. (S31) integrates into:

$$-2\bar{a}^{-1/2} + 2\bar{a}^{-3/2} - \frac{6}{5}\bar{a}^{-5/2} + \frac{2}{7}\bar{a}^{-7/2} + \frac{32}{35} \equiv f(\bar{a}). \qquad (S32)$$

So, now, the mass-distance relation reads:

$$q = \sqrt{\frac{3}{f'}} \, \tau_{\text{disk}}^{-1/2} \sqrt{f(\bar{a})} \; . \qquad (S33)$$

Input into the condition of validity of the continuous regime, it gives:

$$\begin{aligned}
\Delta &< 2 r_H / r_R \\
\bar{a} - 1 &< 2\bar{a} \left( \frac{q}{3} \right)^{1/3} \\
\left( 1 - \frac{1}{\bar{a}} \right)^6 &< \frac{2^6}{3 f' \tau_{\text{disk}}} f(\bar{a}) \\
\tau_{\text{disk}} &< \frac{2^6}{3 f'} f(\bar{a}) \left( 1 - \frac{1}{\bar{a}} \right)^{-6} \equiv \frac{2^6}{3 f'} h(\bar{a}) \qquad (S34)
\end{aligned}$$

Instead of an expression for $\Delta_c$, we get an implicit equation for $\bar{a}$, that can be solved numerically. But this turns also into an explicit condition on $\tau_{\text{disk}}$.

It is possible to expand $f$ and $h$ to first order in $\Delta$. Actually, $f$ is the integral of

$$\Delta^3 (1+\Delta)^{-9/2} d\Delta = \left( \Delta^3 - \frac{9}{2} \Delta^4 + O(\Delta^5) \right) d\Delta$$

So, $f(\Delta) = \frac{\Delta^4}{4} \left( 1 - 3.6\Delta + O(\Delta^2) \right)$.

$$\left( 1 - \frac{1}{\bar{a}} \right)^{-6} = (\Delta/(1+\Delta))^{-6} = \Delta^{-6}(1 + 6\Delta + O(\Delta^2)) \; .$$

So, the condition Eq. (S34) reads:

$$\begin{aligned}
\tau_{\text{disk}} &< \frac{2^6}{3f'} \frac{\Delta^4}{4} \left( 1 - 3.6\Delta + O(\Delta^2) \right) \\
& \quad \times \Delta^{-6}(1 + 6\Delta + O(\Delta^2)) \\
\frac{3 f' \tau_{\text{disk}}}{2^4} &< \Delta^{-2} \left( 1 + 2.4\Delta + O(\Delta^2) \right) \\
\frac{4}{\sqrt{3 f' \tau_{\text{disk}}}} &> \Delta \left( 1 - 1.2\Delta + O(\Delta^2) \right) \qquad (S35)
\end{aligned}$$

To zeroth order, one recognizes the condition $\Delta < \Delta_c \propto \tau_{\text{disk}}^{-1/2}$ where only the numerical coefficient has slightly changed with respect to the previous case Eq. (S10). The right hand member of Eq. (S35) is an approximation of $(4\,h(\bar{a}))^{-1/2}$; while the zeroth order differs from this exact value by more than 10% for $\Delta > 0.1$, the first order is accurate to less than 10% up to $\Delta < 0.25$. Eq. (S35) is a more accurate definition for the boundary of the continuous regime.

### 9.1  Condition for a never-ending continuous regime at small $\tau_{\text{disk}}$

Interestingly, expanding Eq. (S35) to first order gives:

$$1.2\Delta^2 - \Delta + \frac{4}{\sqrt{3 f' \tau_{\text{disk}}}} > 0 \; , \qquad (S36)$$

which is always true if $1 - 19.2/\sqrt{3 f' \tau_{\text{disk}}} < 0$. So, the continuous regime never ends if

$$\tau_{\text{disk}} < \frac{19.2^2}{3 f'} \approx 50 \qquad (S37)$$

The largest possible $\Delta_c$ from Eq. (S36) is $1/2.4 \approx 0.42$. At this point, the first order differs from the exact value $(4\,h(\bar{a}))^{-1/2}$ by $\sim 30\%$. Thus, the value given above should be considered with care. However, it is true that $h(\bar{a})$ admits a minimum, and that for low enough values of $\tau_{\text{disk}}$, only one satellite forms. The numerical integration with accurate torque (SM 8) confirms that $\tau_{\text{disk}} \approx 50$ is the limit (see Figure 3).

# 10 On the influence of the moon on the disk in the continuous and discrete regimes

One can wonder if the torque exerted by the Moon on the disk ($\Gamma$, given by Eq. (S2), SM 2) will not at some point prevent the disk from spreading, and stop the growth of the Moon. To check this, this torque should be compared to the viscous torque exerted by the tidal disk on its outer edge, given by:

$$\Gamma_\nu = 3\pi\nu\Sigma r_{\rm R}^2 \Omega_{\rm R} \ . \tag{S38}$$

## 10.1 Constant $D$, and small moons

Below, we assume that $D$ and $\Sigma$ are constant, that is we consider the case where the mass of the growing moon is small compared to that of the disk: $q \ll D$. Using the relation between the mass and the distance in the continuous and discrete regime Eq. (S9), one finds

$$\Gamma = 2^{-3/2} 3^{-3/4} q^{1/2} \Sigma r_{\rm R}^4 \Omega_{\rm R}^2 \tau_{\rm disk}^{-3/4} \ .$$

Then, using Eq. (S5) and Eq. (S38), one finds:

$$\begin{aligned}
\frac{\Gamma_\nu}{\Gamma} &= \frac{2^{3/2} 3^{7/4} \pi\, 26\, r_h^{*5}\, G^2\, \Sigma^2}{q^{1/2}\, r_{\rm R}^2\, \Omega_{\rm R}^4\, \tau_{\rm disk}^{-3/4}} \\
&= \frac{2^{5/2} 3^{7/4} 13\, r_h^{*5}\, D^2}{\pi\, q^{1/2}\, \tau_{\rm disk}^{-3/4}} \\
\frac{\Gamma_\nu}{\Gamma} &= \frac{9.67}{q^{1/2}\, \tau_{\rm disk}^{1/4}} \tag{S39}
\end{aligned}$$

The ratio between the two torques decreases as $q$ increases. Using the expression of $q_c$ Eq. (S11), the ratio between the two torques at the end of the continuous regime is $\sim 6.9\sqrt{\tau_{\rm disk}}$, generally much larger than 1. Using the expression of $q_d$ Eq. (S13), the viscous torque is an order of magnitude larger than the gravitational torque during all the discrete regime, provided $\tau_{\rm disk} > 20$.

Using Eq. (S6), the ratio of the two torques is $\sim 21\sqrt{D/q} \approx \sqrt{400\, M_{\rm disk}/M}$, so the viscous torque in the disk always dominates the satellite's torque if the satellite isn't more massive than the disk.

In conclusion, the influence of the moon(s) on the disk can be neglected in the continuous and discrete regimes as long as $q < D$.

## 10.2 Varying $D$, and large moons

In this section, we consider the case where the assumption $q \ll D$ doesn't hold anymore, so that $D$ can't be considered as constant. Assuming that at least $F$ is constant (see SM 7.1.2), the mass-distance relation is given by Eq. (S26), which input into Eq. (S2) gives

$$\Gamma = 2^{-3/2} 3^{-3/4} q^{1/2} \left(1 - \frac{2q}{3D_0}\right)^{-3/2} \Sigma r_{\rm R}^4 \Omega_{\rm R}^2 \tau_{\rm disk}^{-3/4} \ .$$

The viscosity is not any more given by Eq. (S5), but considering that $\tau_{\rm disk} = T_\nu/T_{\rm R}$, one gets $\nu = r_{\rm R}^2/T_{\rm R}\tau_{\rm disk}$, and thus:

$$\frac{\Gamma_\nu}{\Gamma} = \frac{2^{3/2} 3^{7/4} \left(1 - \frac{2q}{3D_0}\right)^{3/2}}{q^{1/2}\, \tau_{\rm disk}^{1/4}} \tag{S40}$$

The numerator is always larger than 3.7. So, the viscous torque always wins, and the continuous or discrete regimes are not interrupted as long as:

$$q < \left(\frac{192}{\tau_{\rm disk}}\right)^{1/2} \tag{S41}$$

For the Earth's Moon, this makes $\tau_{\rm disk} < 1.3 \times 10^6$, or a spreading time shorter than $10^3$ years for the spreading never to be blocked by the Moon's torque on the disk.